\documentstyle[epsf,11pt]{article}
\def\baselinestretch{1.0}
\title{\bf Rescaled range and transition matrix analysis\\
 of DNA sequences
}
\author{  Zu-Guo Yu$^{1,2,3}$ and Guo-Yi Chen$^{2}$ \\
 {\small $^{1}$Department of Mathematics, Xiangtan University, Hunan 411105, P.R. China\footnotemark.}\\
 {\small $^{2}$Institute of Theoretical Physics, Chinese Academy of Sciences},\\
  {\small P.O. Box 2735, Beijing 100080, P. R. China.}\\
 {\small $^{3}$CCAST( World Laboratory), P.O. Box 8370, Beijing 100080, P.R. China.}
  }
 \date{}
\newcommand{\be}{\begin{equation}}
\newcommand{\ee}{\end{equation}}

\begin{document}
\maketitle
 \renewcommand{\thefootnote}{\fnsymbol{footnote}}
 \footnotetext{* This is the corresponding address of the  first author,
          	Email: yuzg@itp.ac.cn } 

\begin{abstract}
In this paper we treat some fractal and statistical features of the 
DNA sequences.
 First, a fractal record model of DNA sequence is proposed by mapping 
DNA sequences to
  integer sequences, followed by $R/S$ analysis of the model
 and computation of the Hurst
  exponents. 
Second, we consider transition 
between the four kinds of bases within DNA.
The transition matrix analysis of DNA sequences shows that some measures
of complexity based on transition
proportion matrix  are of interest.
The main 
  results are: 1) $H_{\mbox{exon}}> H_{\mbox{intron}}$
for virus. But $H_{\mbox{intron}}>H_{\mbox{exon}}$ 
for the 
species which have the shape of cell except for drosophila. 
2) For Virus, E. coli, yeast, drosophila, mouse and human,
   measures  ${\cal H}$ of transition 
 proportion matrix of exon is larger than that of intron, and measures
$\lambda$, ${\cal D}, {\cal C}, \widetilde{D}$ and $\widetilde{C}$ of transition 
 proportion matrix of intron  are larger than that of exon. 
3) Regarding the evolution, we find that when the species goes higher in
grade,  the measures
  ${\cal D}$, ${\cal C}$, $\widetilde{D}$ and $\widetilde{C}$ of exon 
become larger,
  the measure ${\cal H}$ of exon becomes lesser except for yeast. 
Hence for species of higher grade, the transition rate among the four kinds
of bases goes further from the equilibrium.
 \end{abstract}

 {\bf Key words}: DNA sequence, functional region, $R/S$ analysis, 
transition proportion matrix, measure of complexity.
\vskip 0.2cm

{\bf PACS} numbers: 87.10 +e

\section{Introduction}
 \ \ In the past decade or so there has been a ground swell of interest in
unraveling the mysteries of DNA. In order to distinguish coding regions from 
non-coding ones, many approaches have been proposed. 
First, investigation into nucleotide
correlation is of special importance. In recent years many authors have 
discussed 
the correlation properties of nucleotides in DNA sequences$^{[1-9]}$. C.K. 
Peng {\it et al}$^{[4]}$, using the
one-dimensional DNA walk 
model found that there exists long-range correlation in non-coding 
regions 
but not
in coding regions. 
Second, the linguistic approach. DNA sequence can be regarded,
at a number of levels,
 as analogous  to mechanisms 
of processing other 
kind of languages, such as natural languages and computer languages$^{[10]}$.
 R.N. Mantegna {\it et al} also studied the linguistic feature of non-coding 
DNA sequences$^{[11]}$.
 Third, the
nonlinear scaling method, such as complexity$^{[12]}$ and 
fractal analysis$^{[13-17]}$. 
Recently, we 
investigated 
the correlation dimension and Kolmogorov entropy of DNA sequences using 
time series model$^{[18]}$. 
Our goal is to search for a good measure of complexity which can be used 
to clearly distinguish
different functional regions of DNA sequences and to describe the evolution 
of species.

\par
In this paper, we first map DNA sequence to sequence of integer numbers, 
and treat it 
like a fractal record in time, then apply $R/S$ analysis to calculate 
its Hurst exponent.
 Second. We
analyze DNA sequences with the transition matrix method and calculate some
measures of complexity
 based on their transition proportion matrices.

\par

\section{$R/S$ analysis}
\ \ A DNA sequence may also be regarded as a sequence over the alphabet 
$\{A,C,G,T\}$, which represents the set of the four 
bases from which DNA 
is assembled, namely adenine, cytosine, guanine and thymine. 
For any DNA sequence $s=s_1s_2\cdots s_N$,
 we define a map $f:\ s\mapsto x=\{x_1,x_2,\cdots,x_N\}$,     
where for any $1\le k\le N$,
\be x_k=\left\{\begin{array}{ll} -2,\quad &\mbox{if}\ s_k=A,\\
     -1,\quad &\mbox{if}\ s_k=C,\\
     1,\quad &\mbox{if}\ s_k=G,\\
     2, \quad & \mbox{if}\ s_k=T. \end{array}\right. \ee

\par
According to the definition of $f$, the four bases $\{A,C,G,T\}$ are mapped to
four distinct value. One can also use $\{-2,-1,1,2\}$ to replace $\{A,G,C,T\}$ or
other orders of $A,G,C,T$. our main aim is distinguish $A$ and $G$ from purine,
$C$ and $T$ from pyrimidine. We expect it to reveal more information than
one dimensional DNA walk$^{[4]}$.

{\bf Remark: }William Y. C. Chen and J. D. Louck $^{[19]}$ also use the 
$\{-2,-1,1,2\}$ alphabet for the DNA sequence, instead of $\{A,C,G,T\}$.

Thus we obtain a number sequence $x=\{x_k\}_{k=1}^N$, 
where $x_k\in\{-2,-1,1,2\}$. 
This sequence can be treated as a fractal records in time.
To study such sequences,
Hurst$^{[20]}$ invented a new statistical method --- 
{\it the rescaled range analysis} ($R/S$ analysis), then 
B.~B.~Mandelbrot$^{[21]}$ and J. Feder $^{[22]}$ introduced $R/S$ analysis of 
fractal records in time  
into fractal analysis.
For any fractal records in time $x=\{x_k\}_{k=1}^N$ and 
any $2\le n\le N$, one can define

\be <x>_{n}=\frac{1}{n}\sum_{i=1}^{n}x_i \ee

\be X(i,n)=\sum_{u=1}^i[x_u-<x>_{n}] \ee

\be R(n)=\max_{1\le i\le n}X(i,n)-\min_{1\le i\le n}X(i,n) \ee

\be S(n)=[\frac{1}{n}\sum_{i=1}^{n}(x_i-<x>_{n})^2]^{1/2}. \ee

Hurst found that

\be R(n)/S(n) \ \sim\ (\frac{n}{2})^H. \ee

 $H$ is called {\it Hurst exponent}.

  As $n$ changes from 2 to $N$, we obtain $N-1$ points 
in $\ln(n)$ v.s. $\ln(R(n) / S(n))$
 plane. Then  we can calculate Hurst exponent $H$ of DNA sequence $s$ 
using the 
least-square linear fit. 
As an example, we plot the graph of $R/S$ analysis of an exon segment $s$ 
of mouse' DNA
sequence (bp 1730-- bp 2650 of the record with Accession AF033620 in Genbank) 
in Figure 1.

\begin{figure}
\centerline{\epsfsize=10cm \epsfbox{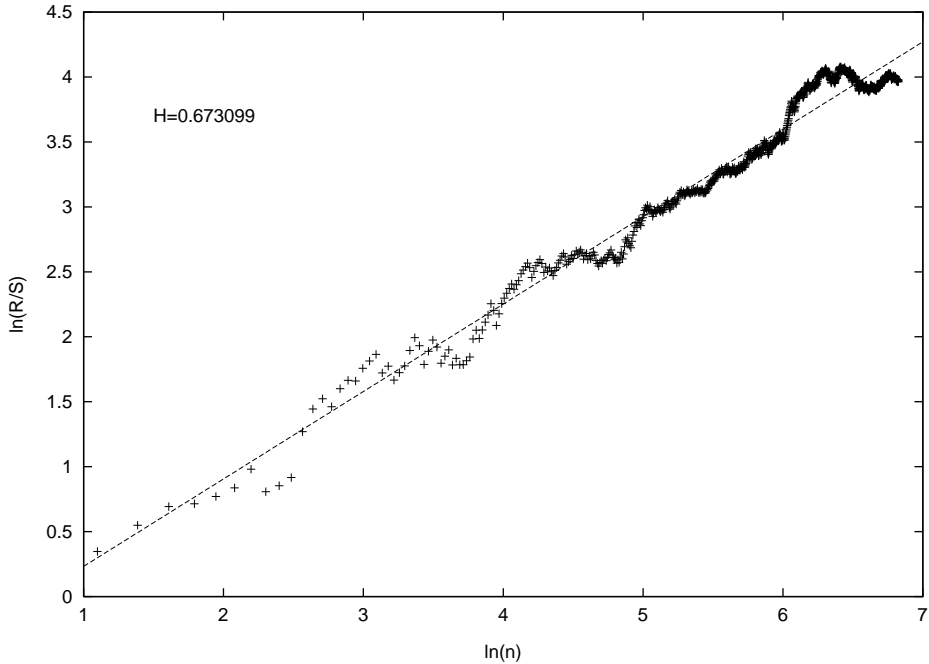}}
\caption{ An example of $R/S$ analysis of DNA sequence}
\end{figure}
\vskip 3mm

  The Hurst exponent is usually used as a measure of complexity. 
From Page 149 of Ref.[22], the trajectory 
of the record is a curve with a fractal dimension $D=2-H$. Hence 
a smaller $H$ means a more complex system.
When applied to fractional Brownian motion, if $H > 1/2$, 
the system is said to be {\em  persistent}, which means that if
for a given time period $t$, the motion is along one direction,
then in the succeeding $t$ time, it's more likely that the motion
will follow the same direction. While for system with $H < 1/2$,
the opposite holds, that is, {\em antipersistent}. But when $H=1/2$,
the system is Brown motion, and is random.

  We randomly choose  17 exons and 34 introns from Virus' genome;
 8 exons and 9 introns from E. coli's;  22 exons and 22 introns from
yeast's;  30 exons and 24 introns
from drosophila's; 37 exons and 31
introns from mouse's; 
 78 exons and 27 introns from Human's( all data from Genbank). 
The Hurst exponent $H$s are calculated for each sequence
and averaged according to both  species category and function,
their relative standard deviations  are also calculated. 
We list the results in Table~1 (we briefly 
write ``relative standard deviation" as ``RSD" in the following tables).

 \begin{table}
\caption{Average and relative standard deviation of $H$ }
\begin{center}
\begin{tabular}{|c|c|c|c|c|c|c|c|}
\hline
  &    &virus &E. coli & yeast & drosophila & mouse & human \\
 \hline
Average&exon & 0.6017 &  0.5991 & 0.6117 &  0.6135 & 0.5746 & 0.5967\\ \cline{2-8}
        &intron & 0.5536 &  0.6482 &  0.6268 &  0.6003 & 0.6017 & 0.6000 \\
 \hline
RSD&  exon & 0.1510 &  0.0790 & 0.1442 & 0.1653 & 0.1446 & 0.1471\\ \cline{2-8}
      &intron & 0.2114 &  0.1265 & 0.1558 & 0.1629 & 0.1795 & 0.1526\\
      \hline
 \end{tabular}
 \end{center}     
\end{table}

  \section{Transition Matrix analysis}
Readers can see the concept of Transition Matrix of a data sequence in the book of
J.C.Davis$^{[23]}$. Here we use this method to study DNA sequences, mainly
on the nature of transitions from one kind of base to another, which
presents useful information of the sequence.

For a given DNA 
sequence $s=s_1s_2\cdots s_N$,   we can construct 
a $4\times 4$ matrix ${\cal A}=(t_{ij})$,
 where  $t_{ij}$ means the number of times a given kind of base being 
succeeded by another in the sequence. ${\cal A}$ is called 
the {\it transition frequency matrix}
 of $s$,
which is a concise way of expressing the incidence of one kind of base
 following another.
 For example, for $s=ATAGCGCATGTACGCGTAGATCATGCTAGCA$, 
the transition frequency matrix is shown below: 

\begin{eqnarray*}
& &\ \ \ \ \ \ \mbox{\bf To}\\
& &\ \ A\ \ T \ \ G\ \ C\\
&\mbox{\bf From}\ \ \begin{array}{l} A\\ T \\ G \\ C\\ \end{array} &\left [
\begin{array}{cccc} 0 & 4 & 3 & 1 \\ 4 & 0 & 2 & 1\\
1 & 2 & 0 & 5 \\
3 & 1 & 2 & 0 \end{array}\right ]. 
\end{eqnarray*}

The tendency for one kind of bases to succeed another can be 
emphasized by converting 
the frequency matrix to
decimal fractions or percentages. 
Therefore, we can construct a matrix ${\cal P}=(P_{ij})$ 
by dividing each element by the grand total 
of all entries in ${\cal A}$. Such a matrix represents the 
relative frequency of all the 
possible types of transitions, and is called the
{\it transition proportion matrix} of $s$.  For the above example,
 the transition proportion matrix is:
\begin{eqnarray*}
& &\ \ \ \ \ \ \ \ \ \ \ \ \ \ \mbox{\bf To}\\
& &\ \ \ \ A\ \ \ \ \ \ T \ \ \ \ \ G\ \ \ \ \ C\\
&\mbox{\bf From}\ \ \begin{array}{l} A\\ T \\ G \\ C\\ \end{array} &\left [
\begin{array}{cccc} 0 & 0.03 & 0.10 & 0.03 \\ 0.03 & 0 & 0.07 & 0.03\\
0.03 & 0.07 & 0 & 0.17 \\
0.10 & 0.03 & 0.10 & 0 \end{array}\right ]. 
\end{eqnarray*}

First, We calculate the maximum real eigenvalue $\lambda$ of 
the transition proportion matrix ${\cal P}$ of
 the DNA sequence. It 
is natural that such a parameter is relevant to the  system's  complexity.

Second, Since $\sum_{i,j=1}^4P_{ij}=1$, $0\le P_{ij}\le 1$, we can view 
$P_{ij}$ as the probability 
of one kind of base to succeed 
another. If we denote $\#\{P_{ij}:\ P_{ij}\neq 0\}=M$ 
be the number of probabilities which is 
not  zero, and rewrite $\{P_{ij}:\ P_{ij}\neq 0\}$ as $\{P_{i}\}_{i=1}^M$. 
Then Shannon's$^{[24]}$ definition of {\it information entropy} applies

\be {\cal H}=-\sum_{i=1}^M P_i\ln P_i. \ee

When $P_i=1/M, i=1,2,\cdots, M$, i.e. the
 case of equilibrium state, the function
 ${\cal H}(P_1,\cdots, P_M)$ reaches its maximum value. 
When $P_i=1$ for some
 $i$ and $P_j=0$ for $j\neq i$, we have ${\cal H}(P_1,\cdots, P_M)=0$.

 There is also a definition of {\it disequilibrium} 
${\cal D}$ $^{[25]}$, used as a measure of "complexity" in $M$-system.

 \be {\cal D}=\sum_{i=1}^M(P_i-\frac{1}{M})^2. \ee

 When $P_i=1/M, i=1,2,\cdots, M$, i.e. the
 case of equilibrium state, the function
 ${\cal D}=0$. When $P_i=1$ for some
 $i$ and $P_j=0$ for $j\neq i$, ${\cal D}$
 gets its maximum value.

 R. Lope-Ruiz {\it et al}$^{[26]}$ proposed another 
statistical measure of complexity ${\cal C}$,
  which is defined as 
 
\be {\cal C}={\cal H}\times {\cal D}. \ee

 Now ${\cal C}=0$ for both the equilibrium state and the case of
$P_i=1$ for some
 $i$ and $P_j=0$ for $j\neq i$.

 We also define two more measures of complexity as follows:
 \be \widetilde{D}=[{\cal D}/(\frac{1}{M}\sum_{i=1}^M P_i^2)]^{1/2} \ee
 \be \widetilde{C}={\cal H}\times \widetilde{D}. \ee

$\widetilde{D}$ means the relative disequilibrium. They are inspired
by ${\cal D}$ and ${\cal C}$, but exhibit better behavior in the computation.

 For DNA sequences chosen in the previous section,
The measures $\lambda$, ${\cal H}$, ${\cal D}$, ${\cal C}$, 
$\widetilde{D}$ and
 $\widetilde{C}$ of complexity  are
 calculated for each sequence
and averaged according to both biological category of species and 
the function. 
In addition, the relative standard deviations of
${\cal H}$, ${\cal D}$, ${\cal C}$, $\widetilde{D}$ and $\widetilde{C}$ 
are also calculated.
 The results are listed in Table~2-7.

 \begin{table}
\caption{Average of the maximum real eigenvalue  $\lambda$ }
\begin{center}
\begin{tabular}{|c|c|c|c|c|c|c|}
 \hline
      &virus & E. coli & yeast & drosophila & mouse & human \\
 \hline
exon & 0.2564   &0.2616   &0.2663   &0.2648  &0.2596  &0.2711\\
 \hline
intron & 0.2913   &0.28835  &0.2980   &0.2839  &0.2752  &0.2720\\
   \hline
 \end{tabular}
 \end{center}     
\end{table}

 \begin{table}
\caption{Average and relative standard deviation of information entropy ${\cal H}$ }
\begin{center}
\begin{tabular}{|c|c|c|c|c|c|c|c|}
 \hline
  &    &virus &  E. coli & yeast & drosophila & mouse & human \\
 \hline
Average& exon & 2.6646   &2.6636   &2.6282   &2.6620  &2.6471  &2.5954\\ \cline{2-8} 
 &     intron &2.5566   &2.5513   &2.5241   &2.5840  &2.5834  &2.5884  \\
 \hline
RSD&exon & 0.0352   &0.0212   &0.0248   &0.0258  &0.0215  &0.0311\\ \cline{2-8} 
&        intron & 0.0770   &0.0268   &0.0401   &0.0398  &0.0372  &0.0339\\
 \hline
 \end{tabular}
 \end{center}     
\end{table}

 \begin{table}
\caption{Average and relative standard deviation of ${\cal D}$ }
\begin{center}
\begin{tabular}{|c|c|c|c|c|c|c|c|}
 \hline
  &    &virus & E. coli & yeast & drosophila & mouse & human \\
 \hline
Average&exon & 0.0121   &0.0123   &0.0172   &0.0137  &0.0146  &0.0211\\ \cline{2-8} 
       &intron & 0.0317   &0.0275   &0.0331   &0.0250  &0.0242  &0.0234 \\
 \hline
RSD&exon & 0.5986   &0.4197   &0.4086   &0.5277  &0.4082  &0.4260\\ \cline{2-8} 
         &intron & 0.7604   &0.2823   &0.4501   &0.5147  &0.5236  &0.5005\\
          \hline
 \end{tabular}
 \end{center}     
\end{table}
  
 \begin{table}
\caption{Average and relative standard deviation of ${\cal C}$ }
\begin{center}
\begin{tabular}{|c|c|c|c|c|c|c|c|}
 \hline
  &    &virus & E. coli & yeast & drosophila & mouse & human \\
 \hline
Average& exon &0.0313   &0.0325   &0.0448   &0.0360  &0.0382  &0.0540\\ \cline{2-8} 
       & intron & 0.0739   &0.0697   &0.0820   &0.0631  &0.0612  &0.0595 \\
 \hline
RSD&  exon & 0.5614   &0.4038   &0.3912   &0.5078  &0.3846  &0.3862\\ \cline{2-8} 
     &intron & 0.7203   &0.2660   &0.4102   &0.4915  &0.4883  &0.4629\\
      \hline
 \end{tabular}
 \end{center}     
\end{table}
  
 \begin{table}
\caption{Average and relative standard deviation of $\widetilde{D}$ }
\begin{center}
\begin{tabular}{|c|c|c|c|c|c|c|c|}
 \hline
  &    &virus &  E. coli & yeast & drosophila & mouse & human \\
 \hline
Average&exon &0.3767   &0.3925   &0.4492   &0.4008  &0.4226  &0.4871 \\ \cline{2-8} 
        &intron & 0.4852   &0.5434   &0.5679   &0.4999  &0.4996  &0.5000 \\
 \hline
RSD& exon & 0.2545    &0.1919   &0.1832   &0.2434  &0.1654  &0.1579\\ \cline{2-8} 
     &intron & 0.3416     &0.1210   &0.1469   &0.2428  &0.2105  &0.1775\\
      \hline
 \end{tabular}
 \end{center}     
\end{table}
  
 \begin{table}
\caption{Average and relative standard deviation of $\widetilde{C}$ }
\begin{center}
\begin{tabular}{|c|c|c|c|c|c|c|c|}
 \hline
  &    &virus  & E. coli & yeast & drosophila & mouse & human \\
 \hline
Average&exon &0.9949     &1.0413   &1.1754   &1.0603  &1.1149  &1.2584\\ \cline{2-8} 
       &intron & 1.2070    &1.3821   &1.4254   &1.2794  &1.2809  &1.2865 \\
\hline
RSD&  exon & 0.2160    &0.1721   &0.1613   &0.2190  &0.1439  &0.1286\\ \cline{2-8} 
     &intron & 0.2722    &0.1002   &0.1105   &0.2122  &0.1774  &0.1435\\
      \hline
 \end{tabular}
 \end{center}     
\end{table}

\section{Conclusions}
  Virus is species which has not the shape of cell. E. coli  
belongs to prokaryote and has the shape of cell. Yeast, drosophila, mouse and human
 belong to eukaryote and also have the shape of cell. From the point of
view of evolution,
virus has lower grade than E. coli; E. coli has lower grade than 
that of yeast
which has lower grade than that of drosophila; drosophila has lower grade than that 
of mouse which
has lower grade than that of human. We use $H_{\mbox{exon}}$ to denote the 
Hurst exponent
of exon, and  similarly for other measures of complexity and 
functional
regions of DNA.

   {\bf  1.} From Table 1, we can see that 
$H_{\mbox{exon}}> H_{\mbox{intron}}$
holds for virus,
but $H_{\mbox{intron}}>H_{\mbox{exon}}$ for the 
species which have the shape of cell except for drosophila. 
The latter means that exons  are more 
complex than introns. 
This result coincides with the conclusion of Ref.[12, 14,18].  
From Table 1 we also find that the Hurst exponent of DNA sequence is 
generally larger than $\frac{1}{2}$. This means that when we use fractional
Brownian motion model to describe DNA sequences, we can say it is a persistent system.
In particular, we can see $H_{\mbox{exon}}$s are different from 1/2 explicitly. It indicates 
that coding regions of DNA is far from random. This is different from the result
of Ref.[4] and coincides with the results of Ref.[14].
But we can not find any trend that coincides with the evolution in Table 1. 

   When we consider the transition of bases in DNA sequence, then 

   {\bf  2}. For Virus, E. coli, yeast, drosophila, mouse and human,
   from Table 3, we can conclude that  measure  ${\cal H}$ of transition 
 proportion matrix of exon  is larger than that of intron, and measures 
 $\lambda$,
 ${\cal D}, {\cal C}, \widetilde{D}$ and $\widetilde{C}$ of transition 
 proportion matrix of intron  are larger than that of exon.

  {\bf 3.} Regarding the evolution, 
we find that as the grade goes higher,  measures
  ${\cal D}$, ${\cal C}$, $\widetilde{D}$ and $\widetilde{C}$ 
of exon become larger,
  the measure ${\cal H}$ of exon becomes lesser except for
yeast. 
Hence for exon of species of higher grade, the transition statistics of the
four kinds of bases goes further from equilibrium.  

   From the above tables, one can find the information entropy ${\cal H}$
 has the less relative standard deviation than other measures of complexity.
  
  {\bf 4.} From the  previous discussions, 
we find that measure
  ${\cal H}$ is a good measure of complexity which can be 
used to clearly distinguish
different functional regions of DNA sequences and to describe 
the evolution of species.

\section*{ACKNOWLEDGMENTS}

{~~~~}The authors would like to express their gratitude toward 
Prof. Bai-lin Hao
for introduction into this field, useful discussions and encouragement. And to 
Prof. Wei-Mou Zheng, Dr. Zuo-Bing Wu and Yang Zhang for many helpful discussions. 
This project was partially supported by China postdoctoral Science Fundation
No. 98B632.

\end{document}